\documentclass[trackchanges, twocolumn]{aastex701}

\usepackage{hyperref}

\usepackage{xcolor}

\begin{document}

\title{\textbf{One Attempt at Building an Inclusive \& Accessible Hybrid Astronomy Conference: FRB 2025 }}

\correspondingauthor{Alice P. Curtin}
\email{alice.curtin@mail.mcgill.ca}

\author{Alice P. Curtin}
\email{alice.curtin@mail.mcgill.ca}
\affiliation{Department of Physics, McGill University, 3600 rue University, Montr\'eal, QC H3A 2T8, Canada}
\affiliation{Trottier Space Institute, McGill University, 3550 rue University, Montr\'eal, QC H3A 2A7, Canada}
\affiliation{Anton Pannekoek Institute for Astronomy, University of Amsterdam, Science Park 904, 1098 XH Amsterdam, The Netherlands}

\author{Reshma Anna-Thomas}
\email{thomas@astron.nl}
\affiliation{Anton Pannekoek Institute for Astronomy, University of Amsterdam, Science Park 904, 1098 XH Amsterdam, The Netherlands}
\affiliation{ASTRON, Netherlands Institute for Radio Astronomy, Oude Hoogeveensedijk 4, 7991 PD Dwingeloo, The Netherlands}

\author{Amanda M. Cook}
\email{amanda.cook@mail.mcgill.ca}
\affiliation{Department of Physics, McGill University, 3600 rue University, Montr\'eal, QC H3A 2T8, Canada}
\affiliation{Trottier Space Institute, McGill University, 3550 rue University, Montr\'eal, QC H3A 2A7, Canada}
\affiliation{Anton Pannekoek Institute for Astronomy, University of Amsterdam, Science Park 904, 1098 XH Amsterdam, The Netherlands}

\author{Carolina Cruz-Vinaccia}
\affiliation{Department of Integrated Studied in Education, McGill University, 3700 McTavish Street, Montr\'eal, QC H3A 1Y2}
\email{carolina.cruzvinaccia@mail.mcgill.ca}

\author{Jason Hessels}
\email{jason.hessels@mcgill.ca}
\affiliation{Department of Physics, McGill University, 3600 rue University, Montr\'eal, QC H3A 2T8, Canada}
\affiliation{Trottier Space Institute, McGill University, 3550 rue University, Montr\'eal, QC H3A 2A7, Canada}
\affiliation{Anton Pannekoek Institute for Astronomy, University of Amsterdam, Science Park 904, 1098 XH Amsterdam, The Netherlands}
\affiliation{ASTRON, Netherlands Institute for Radio Astronomy, Oude Hoogeveensedijk 4, 7991 PD Dwingeloo, The Netherlands}

\author{Robert Main}
\email{robert.main@mcgill.ca}
\affiliation{Department of Physics, McGill University, 3600 rue University, Montr\'eal, QC H3A 2T8, Canada}
\affiliation{Trottier Space Institute, McGill University, 3550 rue University, Montr\'eal, QC H3A 2A7, Canada}

\author{Inés Pastor Marazuela}
\email{ines.pastor.marazuela@gmail.com}
\affiliation{Anton Pannekoek Institute for Astronomy, University of Amsterdam, Science Park 904, 1098 XH Amsterdam, The Netherlands}
\affiliation{ASTRON, Netherlands Institute for Radio Astronomy, Oude Hoogeveensedijk 4, 7991 PD Dwingeloo, The Netherlands}

\author{Lauren Rhodes}
\email{lauren.rhodes@mail.mcgill.ca}
\affiliation{Department of Physics, McGill University, 3600 rue University, Montr\'eal, QC H3A 2T8, Canada}
\affiliation{Trottier Space Institute, McGill University, 3550 rue University, Montr\'eal, QC H3A 2A7, Canada}

\author{Vishwangi Shah}
\email{vishwangi.shah@mail.mcgill.ca}
\affiliation{Department of Physics, McGill University, 3600 rue University, Montr\'eal, QC H3A 2T8, Canada}
\affiliation{Trottier Space Institute, McGill University, 3550 rue University, Montr\'eal, QC H3A 2A7, Canada}

\collaboration{9}{On Behalf of the FRB 2025 Scientific \& Local Organizing Committees}

\begin{abstract}

The rapid expansion of the Fast Radio Burst (FRB) field has been accompanied by a simultaneous growth of FRB conferences. While these meetings are essential for interacting with other researchers and establishing collaborations, many remain only accessible to those with substantial travel funding, flexible schedules, or geographical proximity. This introduces barriers that predominantly affect early career researchers (ECRs) and people from under-resourced regions, limiting the growth, diversity, and sustainability of the community. To address these issues, the FRB~2025 conference, held in Montr\'eal in July 2025 with over 200 participants, was designed to prioritize inclusivity and accessibility alongside scientific excellence.
In this work, we describe how we implemented these goals, including:  organizing committees spearheaded by ECRs, a fully hybrid format including YouTube livestreams, low registration fees, a pedagogical day at the beginning of the conference, local vegetarian catering, and the implementation of flash-talks instead of posters. 
From a post-conference survey of participants, we were able to assess the effectiveness of our initiatives. Notably, we received very positive feedback from the online participants, which amounted to roughly half of the attendees, especially regarding the livestreams and talk recordings. The pedagogical day was also greatly appreciated. The low registration fees naturally led to challenges, in particular with the audio-visual management, and although areas for improvement were noted, such as poster sessions and support for attendees requiring visas, the conference was generally viewed as a success. 
Our experience demonstrates that highly accessible, hybrid conferences are possible within modest budgets ($\$20$k CAD), and we outline recommendations for future conferences, both in the FRB field and in other domains. 

\end{abstract}


\section{Introduction} 

In a given year, there are hundreds of possible conferences and schools for researchers to attend within the field of astronomy. Major astronomical societies typically host one or two large annual meetings (e.g., in the U.S., the AAS Winter and Summer meetings), complemented by numerous smaller, field-specific gatherings. In 2019 alone, approximately 362 such astronomy events were held worldwide \citep{gokus_astronomys_2024}. The large number of possible conferences introduces several difficulties, especially for early career researchers (ECRs)\footnote{While not a strict definition, we use Early Career Researcher throughout to refer to students, post-docs, and researchers within the first few years of their faculty appointment.}. There is significant pressure on students to attend multiple conferences per year to gain visibility within a community and to add content to their CV. Yet, attending conferences is expensive, requires a substantial time investment, can be difficult with respect to obtaining visas, is challenging to balance alongside other personal commitments, and has a high carbon footprint. Combined, these factors can make it difficult to attend more than one per year. Notably, few conferences are fully hybrid\footnote{Here, and within this document, we define a `hybrid' conference as one in which all of the presentations and related-content are equally available to those in-person and those online.} and so attending in-person is often the only way to interact with participants and the relevant content. 

There have been significant efforts across a variety of disciplines to work towards more accessible, inclusive conferences \citep{Joo2022InclusiveConference, kun2023designinginclusiveengaginghybrid, harrison2024bringingafricaneuropean, 2025NatAs...9...11M, 2025NatAs...9....6M}. Certain techniques and takeaways from these efforts include broad advertising to encourage diverse representation, offering financial and logistical support, blind abstract selection, thoughtful scientific programs, and operating as a fully hybrid conference. \citet{2022NatAs...6.1105M} and \citet{2025NatAs...9....6M} emphasized that hybrid conferences are possible on small operating budgets, and do greatly increase the accessibility of these events. On the other hand, the American Astronomical Society found that despite hosting three fully hybrid meetings, the participation online was $<2\%$. Thus, they decided to move away from hybrid meetings, with AAS 243 in January 2024 having a limited online component\footnote{\url{https://aas.org/posts/news/2023/08/future-virtual-participation-aas-meetings}}. 

In this article, we discuss how we addressed some of the current challenges with conferences when organizing FRB 2025, a hybrid astronomy conference held in Montr\'eal in July 2025 with over 200 participants. In Section~\ref{Sec:ConferenceHistory}, we provide background on the yearly `FRB' (fast radio burst) conferences. In Section~\ref{Sec:Setting Priorities}, we describe our initial priorities for the conference, and then in Section~\ref{Sec:Structuring FRB 2025} we discuss the specific actions we took to these ends. In Section~\ref{Sec:Improvements for the Future}, we discuss the impact of our efforts through both qualitative and quantitative feedback from a post-meeting participant survey. We conclude in Section \ref{sec:LookingFuture} with a summary of the lessons we learned and challenges for which we have yet to find a solution.

\section{Conference History}
\label{Sec:ConferenceHistory}

The field of FRB research is rapidly growing and is relatively young compared to many other astronomical sub-fields. In 2024 alone, there were over 250 peer-reviewed publications on FRBs, a substantial jump from the approximately 30 published in 2010\footnote{Compiled using the ADS library: \url{https://ui.adsabs.harvard.edu}}. While the first FRB, a millisecond duration burst of energy from extragalactic distances,
was discovered in 2007 \citep{lorimer_bright_2007}, the first major FRB meeting was not held until 2017. As a result of the field's rapid expansion, these meetings quickly developed into an annual conference series. Now a community staple, the `FRB 20XX' meetings are considered a top priority for many FRB scientists with typically $\sim$100 -- 150 participants.

Reflecting the international nature of the FRB community,
the FRB meetings have been held in the USA (2017), Australia (2018), The Netherlands (2019), South Korea (2022), and Thailand (2024). These locations were popular owing to the relative ease of travel from many countries, particularly Australian and China-based FRB scientists, and the ease of obtaining visas for these locations. Three iterations (2020, 2021, 2023) were also held online, due to both COVID restrictions as well as political constraints.

In 2024, we (Amanda Cook and Alice Curtin) were senior graduate students at the University of Toronto and McGill University, respectively, working on FRBs. Having just booked our flights to Thailand for FRB 2024, we started reflecting on the recent pattern of FRB conference locations. While the meetings had been highly successful, the travel required for the meetings had been challenging for those in North America.

Although Canada is a high-income country and researchers there enjoy many advantages, in some Canadian research groups the grant sizes still meant that it was not possible for all graduate students to attend every international conference. Hence, preference to conference travel was often given to students who were selected to give talks or more senior students. We also noticed that many conferences we wished to attend were not fully hybrid, and so if we were to only attend online, we'd miss out on many aspects of the conference.

Additionally, the carbon emissions associated with each intercontinental trip was enormous. For example, the trip from Montréal to South Korea was approximately 1.5 tCO$_2$e, roughly half of the suggested CO$_2$ budget for each of us for the year.

Some of the previous FRB conferences had taken steps to address some of these concerns. For example, at FRB 2024, the SOC opted to give all who submitted abstracts at least a 3+2 minute talk. This enabled many scientists to attend who had travel funding only available to them when they were giving presentations. Additionally, with three iterations online, this greatly reduced the carbon footprint of the conferences \citep{Burtscher_2020}. On the other hand, there had yet to be a fully hybrid conference.

As we reflected on previous meetings, we felt it might be a good time to host an iteration of the conference series in Canada. At the same time, we decided to commit to making the conference as financially accessible and inclusive as possible, realizing that no location would be convenient for every possible FRB scientist. Thus, we forged ahead with FRB~2025, securing approval from the unofficial annual conference steering committee and starting the planning process for a hybrid conference in Montr\'eal.

\section{Setting Priorities for an Accessible \& Inclusive Conference} 
\label{Sec:Setting Priorities}

We wanted FRB~2025 to be inclusive, financially accessible, and transparent. With this in mind, we found ourselves focusing on five specific areas for consideration:

\begin{itemize}
    \item \textbf{Promoting the participation of ECRs and providing equal footing for all within the field}: The FRB field is still rapidly growing, with large numbers of new researchers joining each year. For example, from 2021 to 2022, authors from 41 countries had published refereed papers on FRBs while from 2021 to 2025 this number rose to 58 countries\footnote{Compiled using the ADS library:  \url{https://ui.adsabs.harvard.edu}.}.  New researchers may not be well known in the field (relevant for talk selection if anonymous reviewing is not used), may not be familiar with specific FRB terminology, and may not be able to afford high travel, accommodation, and registration costs.
    \item \textbf{Making content available beyond the meeting: } Many attending online are located in different time zones, making watching in real-time difficult.  Participants may wish to re-watch talks to understand the intricacies of the work, and these presentations can serve as useful guides for future students. Additionally, the ability to watch at different times is often essential for those with other personal commitments e.g., care-giving responsibilities, teaching, etc. 
    \item \textbf{Reducing bias in the talk selection process}: It is not always transparent how organizing committees are chosen nor how decisions within them are made. This can lead to apparent biases in final talk selection statistics. 
    \item \textbf{Reducing the climate impact of the conference}:  Fossil fuel burning and climate change pose substantial risks to humanity \citep{watts20192019,parmesan2022climate,Grant:2025aa}, with climate tipping points looming ominously close \citep{armstrong2022exceeding,lenton2025global}. Travel to international conferences has a huge carbon footprint. The average worldwide emission per participant per astronomy conference is 1.1 $\pm$ 0.6 tCO$_2$e, approximately half of the recommended annual CO$_2$ budget per person \citep{gokus_astronomys_2024}.  In contrast, fully-online conferences can reduce the total carbon footprint by over 90$\%$ \citep{Burtscher_2020, 2021NatCo..12.7324T}, demonstrating the significant benefit of having online or hybrid conferences.
    \item \textbf{Increasing the financial accessibility of the conference}: Most conferences cost hundreds of dollars, or more, in registration fees. When combined with hotel and flight costs, international conference attendance can easily cost upwards of \$2000~CAD. As an example, when flying from Montr\'eal, FRB~2024, in Thailand, cost a total $\sim$\$4000~CAD. The steep price tag limits who is able to attend in person. 
\end{itemize}

\section{Structuring FRB 2025}
\label{Sec:Structuring FRB 2025}
With these priorities in mind, we designed FRB~2025. Below, we discuss some of our specific efforts. 

\subsection{Composition of the Local and Scientific Organizing Committees}
To promote and highlight ECRs in the FRB community, FRB~2025 was intentionally designed with strong ECR leadership and participation. In addition to having two ECRs as co-Chairs of the conference, both the Local Organizing Committee (LOC) and the Scientific Organizing Committee (SOC) were predominantly composed of ECRs, supported by a small number of senior professors for guidance. When forming the SOC, we ensured broad representation from all major FRB collaborations and instruments across continents (e.g., CHIME, MeerKAT, F4,  Realfast, EVN, LOFAR, DSA, CRAFT, GMRT, FAST) as well as diverse gender, racial, and national representation. 

The LOC was similarly composed mostly of ECRs. Throughout the decision-making process, the LOC sought guidance from the Canadian Astronomical Society (CASCA) Sustainability Committee to ensure that best practices were incorporated into the planning and execution of the conference.

\subsection{Abstract Selection Process}
To reduce bias in the abstract review process, all submissions were evaluated through a dual-anonymous review system, with both the SOC chairs as well as the individual SOC members indicating conflicts of interest \textit{prior} to reviewing any abstracts. The abstract review criteria were not shared with participants beforehand. However, for transparency, we detail the criteria used in our evaluation: (1) scientific novelty, (2) suitability for a flash talk instead of a full talk, (3) the reviewer’s confidence in providing a fair and expert assessment, (4) whether similar work had been presented at FRB~2024 and (5) alternate flags, like plagiarism or similarity to an invited talk. While talks were initially ranked based on scientific excellence, our final selection process ensured balanced representation across genders, countries, seniority, and research collaborations. To avoid self-promotion or bias, SOC members were not permitted to give scientific talks.

\subsection{Pedagogical Day}

An intention we had for this conference was to make the scientific content widely accessible across the FRB community. The rapid expansion of the FRB field has brought together astronomers from diverse backgrounds, such as pulsar and transient astronomy, cosmology, theory, instrumentation, and data science, as well as a number of ECRs. While certain FRB concepts and terminology may be well known in one specific FRB sub-field, they may not be easily understood or recognized by all. Thus, to establish a baseline of knowledge for all participants and to promote their engagement during the conference, we structured the first day as a \textit{pedagogical day}, providing an overview of key notions and the current state of the field.

The FRB~2025 pedagogical day was open to the entire community, from newcomers and cross-disciplinary scientists to senior researchers. It focused on in-depth, lecture style presentations rather than presenting specific research project results or focusing on training skills. Many large conferences similarly offer one or multi-day educational workshops right before the meeting \citep{bergeron_guide_2019, al-shammari_early_2025, zhao_63rd_2025, arase_63rd_2025}.  For example, the first day of the CASCA Annual General Meeting is a graduate student day and the regular Pulsar Timing Array (PTA) meetings often have a full `student week' ahead of the main conference. However, these programs are often solely meant for undergraduate and graduate students, whereas the FRB~2025 pedagogical day targeted, and was attended by, a broader audience.

The idea of a dedicated pedagogical day for FRB~2025 emerged early in the planning, with the name chosen to emphasize accessibility without implying a student-only audience. We designed the program to balance breadth and depth, with speakers representing diverse collaborations, career levels, and expertise within and beyond the FRB field. The speakers were contacted in advance with guidance on the intended scope and level of their presentations to ensure coherence and complementarity of the topics\footnote{We did not provide a specific structure for these presentations, but instead gave each invited speaker a fairly narrow topic as well as a target complexity level for the material.}.

The pedagogical day included four sessions totalling five and a half hours of talks. The first three sessions consisted of 30-minute slots, and covered a broad range of foundational topics: transient astronomy and timing statistics, very-long-baseline interferometry, propagation effects, FRB emission mechanisms, and cosmological and circumgalactic medium applications. The final session consisted of 15-minute presentations on different surveys and instruments that have advanced the field.

\subsection{Hybrid Conference}

\subsubsection{Audio and Visual}
To both increase the availability of FRB 2025 content and reduce the climate impact, the conference was fully hybrid. All sessions were streamed via Zoom, with slides being shared from a dedicated laptop at the front of the auditorium and shown on a projector for the in-person attendees. Speakers were requested to upload their slides at the beginning of their presentation day. The auditorium had a set of speakers alongside the wall, attached to a headset microphone at the front for the presenter. 
The audio input to the computer was captured through two separate microphones -- the internal microphone of the presenting laptop, and a `Catchbox'\footnote{\url{https://catchbox.com}}, a portable padded microphone meant to be easily thrown, as an alternative to passing a microphone during a question period. 

Handling two independent audio inputs was not natively supported in Zoom, so we had to create a custom setup, which we did using the audio software `BlackHole'\footnote{\url{https://www.blackhole.audio}}.  The computer's output audio was connected to the auditorium speakers so that the voices of online participants could be broadcast through the room.  Conversely, the speaker, as well as anyone asking a question through the Catchbox, would be transmitted online.  Something that we did not consider was that the Catchbox audio was not broadcast to the auditorium speakers (discussed further in Section~\ref{Sec:FeedbackAudioAv}). 

\subsubsection{YouTube}
\label{sec:youtube}
To ensure the FRB 2025 content would be recorded and accessible after the meeting, we broadcasted the Zoom stream live to YouTube with a $\sim$30\,s delay.
As an unforeseen benefit, YouTube automatic closed captioning was enabled by default, allowing a real-time transcription in many different languages; several attendees commented that they followed remotely in their native language in real time. We saved recordings of the sessions from Zoom as a backup, but found them unnecessary as the full-day YouTube broadcasts were immediately available from the recordings upon concluding each day and were of similar resolution to the Zoom recordings. 

The almost instantaneous Youtube recordings meant that online participants could watch at times convenient for them. These recordings also serve as time capsules of the field and can be found on the FRB Youtube channel\footnote{\url{https://www.youtube.com/playlist?list=PLM64DlS0XW3ZwpNmkY1RHvauerpu0srfb}}. In addition to the FRB 2025 talks, the FRB 2021 talks are also accessible under the same Youtube channel and there are currently efforts to upload other FRB conference recordings as well. 

\subsection{Encouraging Online Participation}
We implemented three strategies to encourage online participation and to allow remote participants to actively engage. The first was a dedicated, in-room LOC member to act as the representative of those online. When an online participant would raise their virtual hand to ask a question, we had the designated LOC member raise a large foam finger to alert the session chair that there was an online question. We then alternated between the online and in-person participant questions. 

The second strategy we employed was using our dedicated Slack space. For many years now, the annual FRB conference series has used a single, paid Slack space that is available year-round (the cost generously split between a few senior faculty who sit on the unofficial steering committee for the conference series). The Slack space is used to advertise job opportunities and conferences as well as simulate discussions both during and between conferences. During FRB 2025, we used Slack as an alternative avenue for asking questions. This worked well for those watching asynchronously and for periods where there were more questions than time allowed. Questions were encouraged to happen on an open Slack channel so that other attendees could see the question and answer, and often times even jump in with their own thoughts or follow-up questions. 

The final strategy we used was a rotating slide deck to introduce remote participants; online participants could opt to have a slide inserted with a picture and biographical information. This information was gathered via a Google form, where participants could opt to input their name and then any desired combination of: their affiliation, where they were attending the conference from, their photo, an email address, and a brief description of their research interests. We then played this slide deck during the intermissions (lunch and coffee) on the main projection screen (which was shared on Zoom) via the auto advance and loop methods of Google slides. Roughly 35\% of the online attendees opted to include themselves on these slides, and this number increased from about 10\% during the first break they were shown. 

\subsection{Poster Sessions}
We decided against offering a poster session for FRB 2025 as poster sessions are notoriously difficult to arrange for fully hybrid conferences. For conferences that are solely online, this is a bit easier as apps such as \textit{Gather Town}\footnote{\url{https://www.gather.town}} allow participants to walk around (virtually) to the different posters and interact with the speakers. However, using something similar with half in-person and half online participants is much harder, particularly in terms of getting in-person participants to meaningfully interact as well as finding ways to reduce feedback between computers. 

Instead, similar to FRB 2024, we offered a small selection of `flash talks': three-minute talks with two minutes for questions. This format allowed both in-person and online participants to give flash talks, and ensured equal accessibility for everyone. However, we were only able to accommodate 15 flash talks, significantly less than the total number of people who did not receive full talks.

\subsection{Reducing the In-Person Footprint}

In addition to the hybrid nature of the conference, we also attempted to mitigate the environmental impact of in-person attendees through two measures. 
The first was to limit single-use products that arise as a result of providing lunch and coffee breaks through disposable cutlery, plates and coffee/water cups. Given the size and location of the conference, it was not possible to completely avoid these products because there was no access to sufficiently large dish-washing facilities. However, we removed the need for single-use plastic cups by providing every conference attendee with a `\textit{Nalgene}' water bottle. \textit{Nalgene}'s are reusable hard-wearing water bottles made of recycled plastic and don't let harmful chemicals leak into the water. Furthermore, we used the Nalgenes as a memento for the conference by having them personalized with the conference logo commissioned by a local  graduate student. 
Additionally, for all products that were single-use, we ensured they were made from recycled products and were recyclable or biodegradable. 

The second measure was through the food we provided. The biggest impact we could make here was only serving vegetarian food. Compared to providing meat options, vegetarian food has dramatically lower CO\textsubscript{2} and methane emissions. While still a significantly smaller carbon footprint than air travel, we estimate that if we had served chicken (or red meat) for lunch each day of the conference, this would have been an additional half a ton (or ten tons) of CO$_2$ emission.

\subsection{Encouraging Early Career \& Under-Resourced Country Participation}

With SOC and LOC members representing a diverse set of countries and collaborations, we were able to widely advertise the conference on over 20 Slack spaces, list servers, and other announcement platforms. This ensured that the conference was promoted worldwide and among various astronomy communities. 

We also kept our registration fees as low as possible, charging \$300~CAD for in-person faculty, \$150~CAD for all other in-person participants, and \$30~CAD for online participants. Moreover, the SOC reviewed fee waiver requests and granted full waivers to the $\sim$70 participants (largely ECRs) who requested them. In addition to the conference itself, the conference dinner fee was very low (\$50~CAD), ensuring broad participation from in-person participants. 

During the conference, most sessions were chaired by ECRs, providing them with visibility, experience, and leadership opportunities within the FRB community. 

\section{FRB~2025 Feedback}
\label{Sec:Improvements for the Future}

To help us reflect on FRB~2025, we conducted a survey of participants within two months of the conference. The survey garnered a total of 60 responses (out of $\sim$200 conference attendees\footnote{There were $\sim$100 in-person and 100 online attendees.}). The responses were split approximately half and half between online and in-person attendees, and consisted of 40\% graduate student responses, 17\% postdoctoral scholars, 31\% staff and faculty, and 12\% undergraduates. Below, we discuss some of the results from that survey as well as some of our own takeaways from the conference. 

\subsection{Online Attendance}

At its peak, there were approximately 100 people attending FRB~2025 remotely through Zoom. This was comparable to the number of in-person participants, and demonstrated the large demand for the online option. Feedback from the survey showed that most participants wished they could have attended in-person, but were unable to largely due to limited access to travel funding as well as other commitments/responsibilities, consistent with our predictions for some of the biggest barriers in being able to attend solely in-person conferences. Thus, despite our efforts to greatly reduce the total cost of the conference through low registration fees, it is obvious that financial barriers continue to play one of the largest roles in online attendance. Contrary to what we expected, environmental factors played a very small role in the decision of whether or not to attend in-person. 

While most online participants watched the conference in real-time (68\%), the rest watched either later that day or after the conference finished, emphasizing the need for conference recordings as well as near-immediate uploading of those recordings. Most noted the time difference as playing the largest role in their ability to participate and watch in real-time.

\subsection{Pedagogical Day}
\begin{figure}
    \centering
    \includegraphics[width=0.7\linewidth]{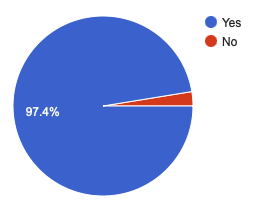}
    \caption{Survey responses to whether future FRB conferences should similarly have a pedagogical day.}
    \label{fig:PedegogicalDay}
\end{figure}

\begin{figure}
    \centering
    \includegraphics[width=0.92\linewidth]{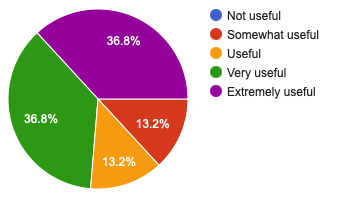}
    \includegraphics[width=0.92\linewidth]{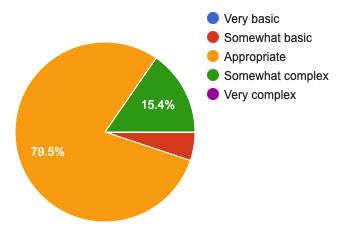}
    \caption{\textbf{Top}: Survey responses as to the usefulness of the pedagogical day content. \textbf{Bottom:} Same as above, but for the complexity of the content.}
    \label{fig:ContentUsefulComplexity}
\end{figure}

This was the first FRB conference to start with a full pedagogical day. As seen in Figure~\ref{fig:PedegogicalDay}, a remarkable 97\% of survey responders said that future FRB conferences should similarly hold a pedagogical day. An overwhelming majority of participants found the content useful and the complexity level appropriate (see Figure~\ref{fig:ContentUsefulComplexity}).  
Although we were uncertain of senior researcher participation, the session drew a broad audience, indicating the community’s interest in this format. Interestingly, the audience and online participation numbers were highest during the pedagogical day. However, it remains unclear whether this was correlated with the pedagogical day itself or with it being the first official day of the conference. 

The YouTube recordings (discussed in more detail in Section~\ref{subsec:ReflectionsYoutube}) of this day provide a lasting introduction to the FRB field. An additional benefit was the beautiful slides and movies made by the speakers to illustrate core concepts. A new slack channel was created in the FRB 20XX slack space for FRB animations, which we hope will be broadly used (with proper credit).

In the open response questions, we received a small number of suggestions for the pedagogical day. Some included having a technical theory/analysis session, having a single chair for the day to tie all of the talks together, and having slightly more time for questions. 

The strong, positive feedback on the pedagogical day strongly demonstrates that not only should the FRB field continue to host these pedagogical days, but other fields should similarly consider full days that introduce non-experts to the field.

\subsection{Audio and Visual}
\label{Sec:FeedbackAudioAv}

 We chose not to have a dedicated AV team, although this greatly increased the burden on our local LOC; getting the correct audio setup in the room, and testing reproducibility, took $\sim2\times1$\,hour sessions, and we needed an LOC member in essentially a full-time role through the conference to handle AV. In total, we estimate the LOC (excluding the chairs) spent well over 100 hours preparing for the conference.  Some participants remarked that having a dedicated AV team for the conference could have been beneficial. We agree, but we were constrained by our overall budget (see Table~\ref{Table: Budget}). The total FRB~2025 working budget (not including the conference dinner) was $\sim$\$15,000 CAD, of which only $\sim$\$6,000 CAD was sourced from registration fees. Having a dedicated AV team would have added an additional \$5,000 -- \$10,000 CAD, nearly doubling the entire working budget and at least doubling the registration costs. Thus, choosing to hire this team would have required us to raise (or charge) the low registration costs for many.

 \begin{deluxetable*}{l c}
\tabletypesize{\normalsize}
\tablewidth{0pt}
\tablecaption{Budget for FRB 2025
\label{Table: Budget}}
\tablehead{
    \colhead{Item} & \colhead{Cost (CAD)}
}
\startdata
Catchbox (in-person microphone)         & 1400 \\
ConfTool (registration platform)         & 1400 \\
Room booking & 900 \\
Coffee (5 days, $\times2$/day) & 1200 \\
Food (5 days) & 7300 \\
Cutlery & 700 \\
Merch (Nalgene bottles, buttons) & 2500 \\
Name tags \& signs & 150 \\
Conference dinner & 5600 \\
\hline
\textbf{Total} & 21150 \\
\enddata
\end{deluxetable*}

In terms of sound quality, $>94$\% of in-person participants could hear most of the time. However, only $\sim$40\% could hear all of the time. Those online did not note the same difficulty in hearing. We believe this was largely due to the lack of a speaker system connected to the Catchbox. While the Catchbox nicely transmitted for online participants, it did not transmit to a speaker system for the in-person room. Thus, it was difficult for in-person participants to hear other in-person participant questions. While we continuously asked people in the room to speak up, this was clearly not enough. For future conferences, we need to ensure that in-person participant voices are amplified through speaker systems. 

In the free-form feedback, participants noted that being able to see those online (through a second screen) as well as a more advanced in-person camera (to Zoom to those speaking in-person) would have created a more cohesive connection between in-person and online participants. 

\subsection{YouTube, Posters, \& Content}
\label{subsec:ReflectionsYoutube}
\begin{figure}
    \centering
    \includegraphics[width=\linewidth]{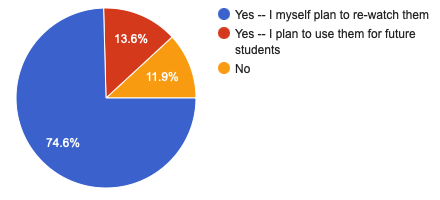}
    \caption{Survey responses to whether participants expect to use the YouTube recordings in the future.}
    \label{fig:Youtube}
\end{figure}

FRB~2025 was the second FRB conference (FRB~2021 also did this) to upload the video recordings to YouTube (see \S \ref{sec:youtube}). Respondents reacted overwhelmingly positively to having the YouTube recordings (89\% said they would use them in the future; see Figure~\ref{fig:Youtube}). To-date, the YouTube recordings have been viewed over 6,000 times. The additional accessibility granted through the YouTube recordings of translating the subtitles in real-time to other languages was also a significant added benefit, and we strongly recommend using something similar in the future.

\begin{figure}
    \centering
    \includegraphics[width=\linewidth]{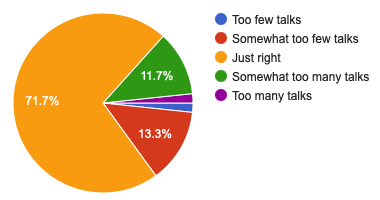}
    \caption{Survey responses to whether the number of talks were appropriate.}
    \label{fig:TalkNumbers}
\end{figure}

FRB~2025 consisted of 37 full talks and 15 flash talks. For most participants (see Figure~\ref{fig:TalkNumbers}), this was the appropriate amount of content. However, we had over 120 abstract submissions, meaning that approximately 3/5ths of all participants did not have any way of promoting their work.  While we did not include a poster session due to the difficulty in making them fully hybrid, many participants (60\%) noted that they would have interacted with them and/or that they would have allowed them to attend in-person. One idea for future, hybrid poster-like sessions could be parallel sessions with three minute flash talks. This could then be accompanied by an in-person poster session where participants could learn more. However, adding the later part would render the conference not fully hybrid. On the other hand, parallel sessions could also prohibit people from attending all the talks they are interested in, and disadvantage sessions with lesser-known presenters in terms of audience size. Another idea, implemented first at FRB 2021, is to allow anyone who submitted an Abstract to upload a short presentation to the YouTube channel. While this still does not provide an in-person opportunity for everyone, it does give each person a chance to disseminate their work.

\subsection{Other Efforts}

We tried to reduce the carbon footprint as much as possible with small efforts. The Nalgene bottles for the conference were received overwhelmingly positively, and no comments were received in-person about a lack of plastic water bottles or plastic cups. Additionally, most participants highly praised the vegetarian food. However, while we offered all participants the option to have specific food ordered for those with dietary restrictions, we received a small number of comments that not having meat was difficult for their diet. 
%In the future, we would advocate to further expand the food choices available to be more inclusive of cultural norms/preferences and religious dietary restrictions. 

We aimed to advertise the conference very broadly to ensure participation from under-represented countries. The hybrid nature of the conference and the low registration fees helped reduce barriers introduced due to limited funding. In the future, in addition to fee waivers, we could offer full travel funding for a couple of participants from under-represented countries. This would ensure more representation and visibility of exceptional researchers from under-resourced communities. 

The announcement of this conference was made several months before the start of the conference to allow for sufficient visa processing times. However, most participants choose to register for the conference and start planning travels only if they are selected to present at the conference. Since the results of abstract submission came out $\sim$ two months before the start of the conference, this significantly reduced the time allowed for visa processing. Thus, almost all participants who required visas were unable to get them on time. This was unlike earlier FRB conferences where visas were relatively easy to obtain. 

\subsection{Time Commitment}

Organizing and preparing for FRB 2025 took a significant amount of time and energy from the LOC. Weeks were spent finding an appropriate venue, acquiring the proper equipment, and testing the audio for in-person and online participants. Additionally, managing the hybrid aspect during the conference, as well as preparing food, coffee, etc., required each LOC member to spend at least a few hours each day of the conference on LOC-related tasks. This created significant post-conference fatigue. While we hope documents like this one can help alleviate that fatigue by providing examples of what worked well vs. what did not, hybrid conferences will still likely require more effort than fully in-person or online efforts. Alternating between different formats (e.g., one year hybrid, another year online) could help with long-term sustainability of such efforts. 

\section{Looking to the Future}
\label{sec:LookingFuture}
Planning and executing inclusive hybrid conferences takes critical thinking and priority setting. We will not get there without significant effort at the community level. This does not mean or require spending lots of money or resources -- a lot can be done with strong Local and Scientific Organizing Committees who are similarly committed to the same principals. FRB~2025 was one such effort to do this. Like all efforts in this domain, FRB~2025 had both its strengths and its weakness. Below, we summarize a few of the things that worked well, as well as things we might try to improve in the future:

What worked well:
\begin{itemize}
    \item \textbf{Pedagogical day}: The pedagogical day had enormous turnout and proved useful to almost all participants. Future conferences should strongly consider similar initiatives that do not just target graduate/undergraduate students, but the wider community.  
    \item \textbf{YouTube livestream}: The YouTube livestream was already integrated as an option within Zoom, and had double benefit of immediate full recordings after each day, and translatable closed captioning of all of the audio. 
    \item \textbf{Hybrid}: The hybrid conference allowed $\times2$ the number of participants to engage with the material of FRB~2025. This was beneficial both for the accessibility of the content as well as the total carbon footprint of the conference.
    \item \textbf{Communication through Slack}: Having a dedicated slack channel provides a place for easy interaction between online and in-person participants. Even better, as the FRB 20XX slack space is a paid slack space that is used year after year, it allows for easy conference promotion, as well as continued, post-conference discussion and collaboration.
    \item \textbf{Catchbox}: The Catchbox enabled clear broadcasting of in-person questions to remote attendees, and facilitated slightly faster question periods compared to mic runners.  
    However, we did need to be careful avoiding laptops and to avoid hitting unaware conference members, and in effect did need LOC `mic-runners' as middle-points for Catchbox throws. 
    \item \textbf{Low cost}: The low registration fees, combined with the relatively low cost of Montréal (compared to other major North American cities) rendered FRB~2025 relatively affordable for a North-American conference. 
\end{itemize}

What can be improved:
\begin{itemize}
    \item \textbf{Visas}: Acquiring visas for Canada is very difficult, and those who required visas were largely not able to obtain them on time. Future conferences could release abstract submission results several months in advance. However, ideally, conferences should be held in places for which a visa is easily obtainable within a couple weeks. 
    \item \textbf{No full funding opportunities}: While we were able to waive the fees for many, this was a relatively small portion of the total cost of attending. Possibly higher registration fees for those who can pay it could offset hotel and/or flight costs for those for which there is no available funding. 
    \item \textbf{Poster sessions}: While we did not include a poster session due to the difficulty in making them fully hybrid, many participants seemed enthusiastic about them. One possible solution could be parallel sessions with shorter flash talks, although there are also downsides to this. 
    \item \textbf{Time commitment from LOC}: While the cost of FRB~2025 was relatively low, this came at the cost of a large burden on the LOC. To maintain the low cost yet minimize the impact on the LOC, we might recommend hiring someone within the university (staff, student) for the week to help with the day-to-day logistics. 
    \item \textbf{Rubric for abstract review}: While we accounted for demographic representation in the abstract review process, we recognize that providing participants with the abstract review rubric beforehand could have enhanced transparency and allowed for a fairer evaluation process. We encourage future conference organizers to consider sharing their rubrics with participants ahead of time.
    \item \textbf{Restricted software applications}: While we discussed the benefits of the Zoom, YouTube live streams, and dedicated Slack channel, we note that these are banned and thus inaccessible in certain countries. 
\end{itemize}

\begin{acknowledgments}
A.P.C. is a Canadian SKA Scientist and is funded by the Government of Canada / est financé par le gouvernement du Canada. 
A.M.C. is a Banting Postdoctoral Research Fellow. 
L.R. acknowledges support from the Trottier Space Institute Fellowship and from the Canada Excellence Research Chair in Transient Astrophysics (CERC-2022-00009). V.S.
is supported by a Fonds de Recherche du Quebec - Nature
et Technologies (FRQNT) Doctoral Research Award. 

FRB 2025 was funded in part by the the Trottier Space Institute and the Canada Excellence Research Chair in Transient Astrophysics (Jason Hessels). We thank the CASCA Sustainability Committee for help with both the conference itself, as well as this document. 
\end{acknowledgments}

\bibliography{bib}{}
\bibliographystyle{aasjournalv7}

\end{document}